\documentclass[prb,showpacs,twocolumn,amsmath,amssymb]{revtex4}
\usepackage{graphicx}
\usepackage{bm}

\def\br{ \bm{r} }
\def\bk{ \bm{k} }
\def\bo{ \bm{0} }
\def\im{ \,\mathrm{Im}\,}
\def\re{ \,\mathrm{Re}\,}
\def\Tr{ \,\mathrm{Tr}\,}
\def\bq{ \bm{q} }

\begin{document}

\title{NMR relaxation time in a clean two-band superconductor}

\author{K. V. Samokhin and B. Mitrovi\'c}

\affiliation{Department of Physics, Brock University,
St.Catharines, Ontario, Canada L2S 3A1}
\date{\today}

\begin{abstract}
We study the spin-lattice relaxation rate of nuclear magnetic
resonance in a two-band superconductor. Both conventional and
unconventional pairing symmetries for an arbitrary band structure
in the clean limit are considered. The importance of the
inter-band interference effects is emphasized. The calculations in
the conventional case with two isotropic gaps are performed using
a two-band generalization of Eliashberg theory.
\end{abstract}

\pacs{74.25.Nf, 74.20.-z}

\maketitle

\section{Introduction}

Although the Fermi surface in most superconductors consists of
more than one sheet, this does not necessarily mean that all those
materials are multi-band superconductors. The true multi-band (in
particular two-band) superconductivity is in fact a rather
uncommon phenomenon characterized by a significant difference in
the order parameter magnitudes in different bands. For this to be
the case, the system has to satisfy some quite stringent
requirements, namely the pairing interactions and/or the densities
of states should vary considerably between the bands and the
inter-band processes, e.g. due to impurity scattering, should be
weak. Although some examples have been known since early
1980s,\cite{Binnig80} the recent swell of interest in this subject
has been largely stimulated by the discovery of two-band
superconductivity in MgB$_2$.\cite{MgB2} Most of the experimental
evidence, see Ref. \onlinecite{MgB2-review} and the references
therein, support the conclusion that there are two distinct
superconducting gaps $\Delta_\pi$ and $\Delta_\sigma$ in this
material, with $\Delta_\sigma/\Delta_\pi\simeq
2.63$.\cite{Golub02} (There are actually four bands crossing the
Fermi level in MgB$_2$, which can be grouped into 2
quasi-two-dimensional $\sigma$-bands and 2 three-dimensional
$\pi$-bands and described by an effective two-band model.) Other
candidates for multi-band superconductivity which have emerged
recently include nickel borocarbides,\cite{CGB98}
NbSe$_2$,\cite{NbSe2} and also the heavy-fermion compounds
CeCoIn$_5$ (Ref. \onlinecite{CeCoIn5}) and CePt$_3$Si (Ref.
\onlinecite{CePtSi}). It seems more likely to find a two-band
superconductivity in unconventional materials, since they are
intrinsically in a clean limit, so at least the gap averaging due
to impurity scattering is not effective.

Theoretically, a two-band generalization of the
Bardeed-Cooper-Schrieffer (BCS) model was introduced independenly
by Suhl, Matthias, Walker,\cite{SMW59} and
Moskalenko.\cite{Moskal59} In subsequent developments, many
aspects of the multi-band model have been studied, including the
thermodynamic and transport properties, the effects of impurities
and strong coupling, \emph{etc}.\cite{GZK67,Chow68,Ent76,SS77}
Surprisingly, little attention has focussed on such an important
characteristic as the spin-lattice relaxation rate $T_1^{-1}$ of
nuclear magnetic resonance (NMR). The measurements of $T_1^{-1}$
probe the properties of the electron subsystem which are local in
real space and therefore extremely non-local in the momentum
space.\cite{Slichter90} In the presence of multiple Fermi-surface
sheets this would give rise to inter-band interference terms in
$T_1^{-1}$, even without any inter-band scattering due to
interactions or impurities. The inter-band terms in $T_1^{-1}$ are
not negligible and can be expected to strongly affect the
temperature dependence of the relaxation rate compared to the
single-band case.\cite{HS59}

The purpose of this article is to calculate the nuclear spin
relaxation rate in a two-band superconductor, for both
conventional and unconventional types of pairing. We focus on
singlet pairing in the absence of impurities, assuming that the
relaxation is dominated by the Fermi contact interaction between
the nucleus and the conduction electrons. The article is organized
as follows. In Sec. \ref{sec: weak coupling}, we develop a general
formalism based on an anisotropic two-band BCS model and show
that, while the resulting expressions in the unconventional case
are well-defined and can be calculated without any additional
complications, in the conventional isotropic case one encounters
divergent integrals. In Sec. \ref{sec: strong coupling}, we single
out the isotropic case for a strong-coupling theory treatment, in
which the divergences are smeared out due to the quasiparticle
lifetime effects. In Sec. \ref{sec: MgB2}, we apply the general
theory to the relaxation rate on the ${}^{25}$Mg site in MgB$_2$
using the realistic strong-coupling parameters.

\section{Weak coupling theory}
\label{sec: weak coupling}

Without the loss of generality, we consider the case of a nuclear
spin $I=1/2$ located at the origin of the crystal lattice. Higher
values of $I$ change only the overall pre-factor in the expression
for the relaxation rate,\cite{Slichter90} which drops out of the
ratio of the relaxation rates in the superconducting and the
normal states. The spin-lattice relaxation rate due to the
hyperfine contact interaction of the nucleus with the band
electrons is given by
\begin{equation}
\label{T1T general}
    R\equiv\frac{1}{T_1T}=-\frac{J^2}{2\pi}\lim\limits_{\omega_0\to
    0}\frac{\im K^R_{+-}(\omega_0)}{\omega_0},
\end{equation}
where $J$ is the hyperfine coupling constant, $\omega_0$ is the
NMR frequency, and $K^R_{+-}(\omega_0)$ is the Fourier transform
of the retarded correlator of the electron spin densities at the
nuclear site:
\begin{equation}
\label{K retard1}
    K_{+-}^R(t)=-i\left\langle\left[S_+(\bo,t),S_-(\bo,0)\right]
    \right\rangle\theta(t).
\end{equation}
Here $S_\pm(\br,t)=e^{iH_et}S_\pm(\br)e^{-iH_et}$, $H_e$ is the
electron Hamiltonian, and
\begin{equation}
\label{Spm}
    S_+(\br)=\psi^\dagger_\uparrow(\br)\psi_\downarrow(\br),\quad
    S_-(\br)=\psi^\dagger_\downarrow(\br)\psi_\uparrow(\br)
\end{equation}
($\hbar=k_B=1$ in our units, and the spin quantization axis is
along the external magnetic field $\bm{H}$). The derivation of Eq.
(\ref{T1T general}) is outlined in Appendix \ref{sec: Appendix A}.
The retarded correlator is obtained by analytical continuation of
the Matsubara time-ordered correlator:
$K^R_{+-}(\omega)=K(\nu_m)|_{i\nu_m\to\omega_0+i0^+}$, with
$\nu_m=2\pi mT$.

We assume that there are two spin-degenerate electron bands in the
crystal (the generalization to an arbitrary number of bands is
straightforward), and neglect the spin-orbit coupling. The
two-band generalization of the BCS Hamiltonian reads
$H_e=H_0+H_{int}$,\cite{SMW59} where
\begin{equation}
\label{H0 weak coupling}
    H_0=\sum\limits_{i,\bk\alpha}\xi_{i,\bk}c^\dagger_{i,\bk\alpha}c_{i,\bk\alpha}
\end{equation}
is the non-interacting part ($i=1,2$ is the band index,
$\alpha=\uparrow,\downarrow$ is the spin projection, and the
chemical potential $\mu$ is included in the band dispersion), and
$H_{int}=H^{(1)}_{int}+H^{(2)}_{int}+H^{(12)}_{int}$ is the
pairing interaction. For anisotropic singlet pairing, we have
\begin{eqnarray}
\label{H pair}
    H^{(i)}_{int}=\frac{1}{2}\sum\limits_{\bk,\bk'}
    V_{ii}(\bk,\bk')c^\dagger_{i,\bk\uparrow}c^\dagger_{i,-\bk\downarrow}
    c_{i,-\bk'\downarrow}c_{i,\bk'\uparrow}\nonumber\\
    H^{(12)}_{int}=\frac{1}{2}\sum\limits_{\bk,\bk'}
    V_{12}(\bk,\bk')c^\dagger_{1,\bk\uparrow}c^\dagger_{1,-\bk\downarrow}
    c_{2,-\bk'\downarrow}c_{2,\bk'\uparrow}\nonumber\\
    +\mathrm{H.c.}
\end{eqnarray}
The Hamiltonians $H^{(1)}_{int}$ and $H^{(2)}_{int}$ describe the
intra-band pairing of electrons, while $H^{(12)}_{int}$ describes
the pair scattering between the bands. The inter-band interactions
of the form
$c^\dagger_{1,\bk\uparrow}c^\dagger_{2,-\bk\downarrow}c_{2,-\bk'\downarrow}c_{1,\bk'\uparrow}$
are suppressed if the band splitting is large compared to all
energy scales relevant to superconductivity. We assume, following
Hebel and Slichter \cite{HS59} that, while the resonance is
observed in a strong field in the normal state, the relaxation
takes place in a uniform superconducting state after switching off
the field.

The pairing symmetry is the same in both bands and is determined
by one of the irreducible representations, $\Gamma$, of the point
group of the crystal. The functions $V_{ij}(\bk,\bk')$ are
non-zero only in a thin energy shell near the Fermi surfaces and
can be represented in a factorized form:
\begin{equation}
    V_{ij}(\bk,\bk')=V_{ij}
    \sum_{a=1}^{d_\Gamma}\varphi_a(\bk)\varphi_a(\bk'),
\end{equation}
where $\varphi_a(\bk)$ are the basis functions, and $d_\Gamma$ is
the dimensionality of $\Gamma$. In the absence of time-reversal
symmetry breaking $\varphi_a$'s can be chosen real. The basis
functions do not have to be the same in both bands, but we neglect
this complication here.

The properties of our superconductor can be described using a
standard field-theoretical formalism in terms of the normal and
anomalous Gor'kov functions:\cite{AGD63}
\begin{eqnarray*}
    &&G_{i,\alpha\beta}(\bk,\tau)=\delta_{\alpha\beta}G_i(\bk,\tau),\\
    &&F_{i,\alpha\beta}(\bk,\tau)=(i\sigma_2)_{\alpha\beta}F_i(\bk,\tau),\\
    &&F^\dagger_{i,\alpha\beta}(\bk,\tau)=(-i\sigma_2)_{\alpha\beta}F^\dagger_i(\bk,\tau),
\end{eqnarray*}
which can be combined into a $2\times 2$ matrix Green's function
\begin{equation}
\label{Nambu GFs}
    \hat G_i(\bk,\tau)=\left(%
\begin{array}{cc}
  G_i(\bk,\tau) & -F_i(\bk,\tau) \\
  -F^\dagger_i(\bk,\tau) & -G_i(-\bk,-\tau) \\
\end{array}%
\right).
\end{equation}
In the mean-field approximation, the interaction Hamiltonian is
reduced to the form
\begin{equation}
\label{H int MF}
    H_{int}=\frac{1}{2}\sum\limits_{i,\bk}\Delta_{i,\bk}
    c^\dagger_{i,\bk\uparrow}c^\dagger_{i,-\bk\downarrow}
    +\mathrm{H.c.},
\end{equation}
where $\Delta_{i,\bk}$ is the superconducting order parameter in
the $i$th band, which can be written as
\begin{equation}
\label{OP expansions}
    \Delta_{i,\bk}=\sum\limits_a
    \eta_{i,a}\varphi_a(\bk),
\end{equation}
with $\eta_{i,a}$ being the order parameter components. Both order
parameters appear at the same critical temperature $T_c$, but have
different temperature dependences, which can be found by solving a
system of $2d_\Gamma$ self-consistency equations for the functions
$\eta_{i,a}(T)$. In the frequency representation, the Green's
functions (\ref{Nambu GFs}) become
\begin{equation}
\label{Nambu GFs omega}
    \hat G_i(\bk,\omega_n)=-\frac{i\omega_n\tau_0+\xi_{i,\bk}\tau_3+
    \hat\Delta_{i,\bk}}{\omega_n^2+\xi_{i,\bk}^2+|\Delta_{i,\bk}|^2},
\end{equation}
where $\tau_i$ are Pauli matrices, $\omega_n=(2n+1)\pi T$, and
\begin{equation}
\label{OP matrix}
    \hat\Delta_{i,\bk}=\left(%
\begin{array}{cc}
  0 & \Delta_{i,\bk} \\
  \Delta^*_{i,\bk} & 0 \\
\end{array}%
\right).
\end{equation}

Now we return to the calculation of the relaxation rate (\ref{T1T
general}). For zero spin-orbit coupling, the spin operators
(\ref{Spm}) can be written in the band representation, using
\begin{equation}
\label{psis}
    \psi_\alpha(\br)=\frac{1}{\sqrt{V}}\sum\limits_{i,\bk}e^{i\bk\br}u_{i,\bk}(\br)c_{i,\bk\alpha},
\end{equation}
where $u_{i,\bk}(\br)$ are the Bloch functions, which are periodic
in the unit cell, and $V$ is the system volume. Inserting these
into Eqs. (\ref{Spm}), one obtains the Matsubara spin correlator
$K(\tau)=-\langle T_\tau S_+(\bo,\tau)S_-(\bo,0)\rangle$, which
can be decoupled in the mean-field approximation, using the
Green's functions (\ref{Nambu GFs}). In the absence of
time-reversal symmetry breaking, one can show that
$u_{i,-\bk}(\bo)=u_{i,\bk}^*(\bo)$. Then, taking the thermodynamic
limit, we have
\begin{equation}
\label{Kpm nu m}
    K(\nu_m)=\frac{1}{2}T\sum\limits_{n}\int\limits_{\bk_{1,2}}
    \Tr[\hat{\cal G}(\bk_1,\omega_n+\nu_m)\hat{\cal
    G}(\bk_2,\omega_n)],
\end{equation}
where
$$
\int\limits_{\bk}(...)=\lim\limits_{V\to\infty}\frac{1}{V}\sum\limits_{\bk}(...)=
\int\frac{d^Dk}{(2\pi)^D}(...),
$$
and
\begin{equation}
\label{cal G def}
    \hat{\cal G}(\bk,\omega_n)=\sum\limits_i|u_{i,\bk}(\bo)|^2\hat G_i(\bk,\omega_n),
\end{equation}
with $\hat G_i(\bk,\omega_n)$ given by Eq. (\ref{Nambu GFs
omega}).

Calculating the matrix traces and the Matsubara sums in Eq.
(\ref{Kpm nu m}) followed by the analytical continuation to real
frequencies, we find that the imaginary part of $K^R_{+-}(\omega)$
is proportional to $\omega$ at $\omega\to 0$. The momentum
integrals are calculated making the usual assumption that
$u_{i,\bk}(\bo)$ and $\Delta_{i,\bk}$ weakly depend on
$\xi_{i,\bk}$ in the vicinity of the Fermi surface (i.e. within
the energy range of the order of $T$). We introduce the local
density of quasiparticle states at $\br=\bo$:
$N(\omega)=N_1(\omega)+N_2(\omega)$ ($\omega>0$), where
\begin{eqnarray}
\label{N ab}
    N_i(\omega)&=&\frac{1}{2}\int\limits_{\bk}|u_{i,\bk}(\bo)|^2\delta(\omega-E_{i,\bk})\nonumber\\
    &=&N_{F,i}\left\langle|u_{i,\bk}(\bo)|^2
    \frac{\omega}{\sqrt{\omega^2-|\Delta_{i,\bk}|^2}}\right\rangle_i,
\end{eqnarray}
where $E_{i,\bk}=\sqrt{\xi^2_{i,\bk}+|\Delta_{i,\bk}|^2}$ is the
Bogoliubov excitation energy in the $i$th band, the angular
brackets stand for the average over the Fermi surface, and
$N_{F,i}=(1/8\pi^3)\int dS_F/|\bm{v}_{F,i}|$ is the density of
states at the Fermi level in the $i$th band. The angular
integration in Eq. (\ref{N ab}) is restricted by the condition
$|\Delta_{i,\bk}|\leq\omega$. We also introduce the function
$M(\omega)=M_1(\omega)+M_2(\omega)$, where
\begin{eqnarray}
\label{M ab}
    M_i(\omega)
    &=&\frac{1}{2}\int\limits_{\bk}\frac{\Delta_{i,\bk}}{E_{i,\bk}}|u_{i,\bk}(\bo)|^2\delta(\omega-E_{i,\bk})
    \nonumber\\
    &=&N_{F,i}\left\langle|u_{i,\bk}(\bo)|^2
    \frac{\Delta_{i,\bk}}{\sqrt{\omega^2-|\Delta_{i,\bk}|^2}}\right\rangle_i.
\end{eqnarray}
Then,
\begin{equation}
\label{T1T result}
    R=J^2\int\limits_0^\infty d\omega
    \left(-\frac{\partial f}{\partial\omega}\right)\left[N^2(\omega)+|M(\omega)|^2
    \right],
\end{equation}
where $f(\omega)=(e^{\omega/T}+1)^{-1}$ is the Fermi function.

For $\Delta_i(\bk)=0$, we have $M(\omega)=0$, and the normal-state
relaxation rate is given by $R_n=J^2N_n^2/2$, where
$N_n=N_{n,1}+N_{n,2}$,
\begin{equation}
\label{Nn ab}
    N_{n,i}=N_{F,i}\left\langle|u_{i,\bk}(\bo)|^2\right\rangle_i.
\end{equation}
Finally, we obtain for the ratio of the NMR relaxation rates in
the superconducting and the normal states
\begin{eqnarray}
\label{T1T final result}
    \frac{R_s}{R_n}=2\int\limits_0^\infty d\omega
    \left(-\frac{\partial f}{\partial\omega}
    \right)\frac{N^2(\omega)+|M(\omega)|^2}{N_n^2}.
\end{eqnarray}
As we pointed out at the beginning of this Section, our result
does not depend on the nuclear spin $I$. The expression (\ref{T1T
final result}) has two notable properties. First, the relaxation
rate is controlled by the local densities of quasiparticle states.
Only in the limit of a single-band isotropic pairing can one
express $R$ in terms of the total density of states and recover
the Hebel-Slichter formula,\cite{HS59} see Sec. \ref{sec: con
pairing} below. Second, the contributions to the spin-lattice
relaxation rate from different bands are not simply additive,
since there are inter-band interference terms in Eq. (\ref{T1T
final result}). These terms are present even in the absence of any
inter-band interactions or impurity scattering and can be traced
back to the local character of the hyperfine coupling
$\bm{I}\bm{S}$, which mixes together the electron states near the
Fermi surface from different bands.

\subsection{Conventional pairing}
\label{sec: con pairing}

The order parameter is ``conventional'' if it transforms according
to the unity representation of the point group
$\Gamma$.\cite{Book} The gap functions $\Delta_{i,\bk}$ can be
isotropic or anisotropic, with $M_s(\omega)\neq 0$ in both cases.

Assuming the isotropic pairing with a uniform order parameter, we
have $\Delta_{i,\bk}=\Delta_i$, where both gap functions can be
chosen real without loss of generality. One can view this as an
extreme case of anisotropic superconductivity on an extended
single sheet of the Fermi surface, in which the gap function is
allowed to take only two values, $\Delta_1$ and $\Delta_2$. The
densities of states become
\begin{eqnarray}
\label{Ns convent}
    N(\omega)=\sum\limits_i
    N_{n,i}\frac{\omega}{\sqrt{\omega^2-\Delta_i^2}},\\
\label{Ms convent}
    M(\omega)=\sum\limits_i
    N_{n,i}\frac{\Delta_i}{\sqrt{\omega^2-\Delta_i^2}}.
\end{eqnarray}
Substituting these expressions in Eq. (\ref{T1T final result}) we
arrive at a logarithmically divergent integral. The origin of this
divergence is the same as in the Hebel-Slichter formula in the
single-band case:\cite{HS59} one has to square the BCS-like
density of quasiparticle states, which is singular at
$E=\Delta_1,\Delta_2$. Allowing for a non-zero NMR frequency
$\omega_0$ yields the relaxation rate which is still much higher
than that observed in experiment.\cite{Tink96}

One can smear out the singularity and cut off the divergence
either by introducing some gap anisotropy,\cite{Masuda62} or by
taking into account the strong-coupling effects, which lead to a
finite lifetime of quasiparticles and therefore to
energy-dependent complex gap functions.\cite{Fibich65} Which
mechanism is more important depends on the material. In Sec.
\ref{sec: strong coupling} below, we adopt the latter point of
view and derive the strong-coupling expression for the relaxation
rate for an isotropic gap.

\subsection{Unconventional pairing}
\label{sec: uncon pairing}

If the order parameter transforms according to a non-unity
representation of the point group, then it follows from the
obvious property of the Bloch functions
$|u_{i,g\bk}(\bo)|^2=|u_{i,\bk}(\bo)|^2$ ($g$ is an arbitrary
element of the point group) that $M(\omega)=0$. Therefore,
\begin{equation}
\label{T1T uncon pairing}
    \frac{R_s}{R_n}=2\int\limits_0^\infty d\omega\left(-\frac{\partial
    f}{\partial\omega}\right)
    \left[\frac{N_1(\omega)+N_2(\omega)}{N_{n,1}+N_{n,2}}\right]^2,
\end{equation}
where $N_i(\omega)$ and $N_{n,i}$ are defined by Eqs. (\ref{N ab})
and (\ref{Nn ab}) respectively. In most cases the integral
converges, because the square-root singularity in the density of
states is smeared out by the intrinsic anisotropy of the gap. The
only exception is an unconventional order parameter with isotropic
gap (e.g. an analog of the $B$-phase of $^3$He in a charged
isotropic superfluid), in which case the integral is again
logarithmically divergent.

Since the inter-band pair scattering $H^{(12)}_{int}$ in Eq.
(\ref{H pair}) induces the order parameters of the same symmetry
in both bands, the low-energy behavior of $N_1(\omega)$ and
$N_2(\omega)$ is characterized by the same power law. If there are
line (point) nodes in the gap, then $N_{1,2}(\omega)\propto\omega$
$(\omega^2)$ at $\omega\to 0$, \cite{Book} and $R\propto T^2$
$(T^4)$ as $T\to 0$.\cite{SU91,Hase96} This behavior has indeed
been observed in most heavy-fermion compounds, for a recent review
see Ref. \onlinecite{Flouq05}.

This picture will change if the gap magnitudes in the bands are
considerably different (as mentioned in the Introduction, there
are indications that this might be the case in such materials as
CeCoIn$_5$ and CePt$_3$Si). For example, if the gap in one band is
much smaller than in the other, then, taking the limit
$\Delta_{2,\bk}\to 0$, one obtains instead of Eq. (\ref{T1T uncon
pairing})
\begin{eqnarray}
\label{T1T different gaps}
    \frac{R_s}{R_n}&=&2\int\limits_0^\infty d\omega\left(-\frac{\partial
    f}{\partial\omega}\right)\nonumber\\
    &&\times\frac{N^2_1(\omega)+2N_1(\omega)N_{n,2}+
    N^2_{n,2}}{(N_{n,1}+N_{n,2})^2}.
\end{eqnarray}
While the last term in the integral contributes to the residual
relaxation rate at $T=0$, it is the second term that controls the
power-law behavior at low $T$: we now have $R={\rm const}+aT$ for
line nodes, and $R={\rm const}+aT^2$ for point nodes.

As an illustration of the above results, let us consider a simple
example of a quasi-two-dimensional two-band superconductor with
circular Fermi surfaces and a $d$-wave gap
$\Delta_{1,\bk}=\Delta_0\cos 2\theta$, which has vertical lines of
nodes. The fraction of the density of states from the electrons in
the unpaired band is $r=N_{n,2}/(N_{n,1}+N_{n,2})$. The
Fermi-surface average in Eq. (\ref{N ab}) can be done
analytically:
\begin{equation}
\label{DoS dwave}
    \frac{N_1(\omega)}{N_{n,1}}=\frac{2}{\pi}\left\{
    \begin{array}{cc}
      \displaystyle xK(x^2) &,\ {\rm if}\ x\leq 1, \\
      \displaystyle K(x^{-2}) &,\ {\rm if}\ x>1,
    \end{array}\right.
\end{equation}
where $x=\omega/\Delta_0$, and $K(x)$ is the complete elliptic
integral of the first kind.\cite{AS65}

\begin{figure}
    \includegraphics[angle=0,width=8cm]{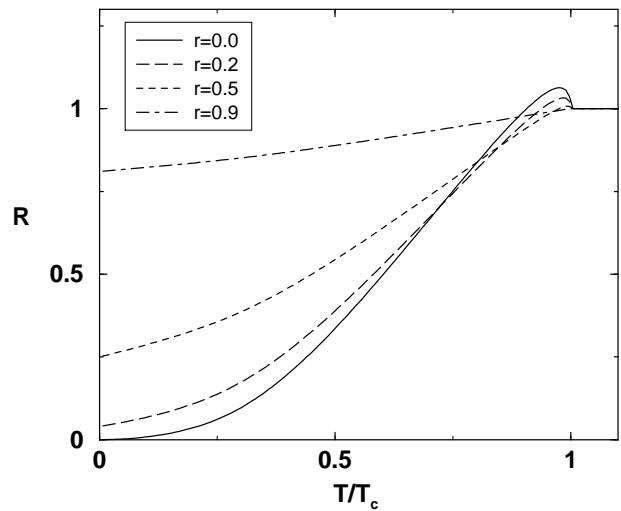}
    \caption{The NMR relaxation rate in a two-band superconductor with
    lines of nodes in one band and a negligible gap in the
    other, for different values of $r=N_{n,2}/(N_{n,1}+N_{n,2})$.}
    \label{fig:dwave}
\end{figure}

In Fig. \ref{fig:dwave}, we show the results of the numerical
calculation of the temperature dependence of the relaxation rate
(\ref{T1T different gaps}) for different values of $r$. Instead of
determining the exact temperature dependence of $\Delta_0$ at all
$T$, which would involve a full numerical solution of the
self-consistency gap equation, we use the approximate expression
$\Delta_0(T)/\Delta_0(0)=\sqrt{1-(T/T_c)^3}$, where
$\Delta_0(0)/k_BT_c=1.30$ (this number is obtained from the
solution of the gap equation at $T=0$). For $r=0$, one recovers
the limit of a single-band $d$-wave superconductor with $R\propto
T^2$ at low $T$ and a small Hebel-Slichter peak immediately below
$T_c$. As $r$ grows, so do both the deviation from the $T^2$
behavior and the residual relaxation rate at $T=0$. One
interesting observation is that even if the density of states is
dominated by the contribution from the unpaired sheet of the Fermi
surface, one still can see an appreciable suppression of the
relaxation rate at low temperatures.

\section{Strong coupling theory}
\label{sec: strong coupling}

In this section we generalize the results of the weak coupling
theory, Sec.~II, to the case of an electron-phonon multi-band
superconductor which could be described by Eliashberg-type
equations.\cite{Ent76,Golub02} To include the self-energy effects
associated with both electron-phonon and screened Coulomb
interaction one replaces Eq. (\ref{Kpm nu m}) with
\begin{eqnarray}
\label{strong_Kpm nu m}
    K(\nu_m)=\frac{1}{2}T\sum\limits_{n}\int\limits_{\bk_{1,2}}\sum_{i,j}
    |u_{i,\bk_1}(\bo)|^2|u_{j,\bk_2}(\bo)|^2\nonumber \\
    \times\Tr[\hat
    G_i(\bk_1,\omega_n)\hat{\Gamma}_{ij}(\bk_1,\bk_2;\omega_n,\nu_m)\nonumber\\
    \times\hat G_j(\bk_2,\omega_n+\nu_m)],
\end{eqnarray}
where $\hat G_i(\bk,\omega_n)$ are given by
\begin{equation}
\label{strong_Nambu GFs omega}
    \hat G_i(\bk,\omega_n)=-\frac{i\omega_n
    Z_{i,\bk}(\omega_n)\tau_0+\xi_{i,\bk}\tau_3+\phi_{i,\bk}(\omega_n)\tau_1}
    {\omega_n^2Z^2_{i,\bk}(\omega_n)+\xi_{i,\bk}^2+\phi_{i,\bk}^2(\omega_n)},
\end{equation}
instead of Eq. (\ref{Nambu GFs omega}). Here $Z_{i,\bk}(\omega_n)$
and $\phi_{i,\bk}(\omega_n)$ are the renormalization function and
the pairing self-energy, respectively, for the $i$th band.

The vertex functions
$\hat{\Gamma}_{ij}(\bk_1,\bk_2;\omega_n,\nu_m)=\hat{\Gamma}_{ij}(\bk,\omega_n;\bq,\nu_m)$
need to be calculated in the conserving approximation consistent
with the approximations used to calculate the electron
self-energies.\cite{ES,Schr,Choi} Since after analytic
continuation $i\nu_m\rightarrow\omega_0+i0^+$ one is interested in
the low-frequency limit, see Eq. (\ref{T1T general}), and the
Migdal's theorem \cite{Migdal,Elias} guarantees that the
electron-phonon contribution to the vertex functions satisfies
$\lim_{\nu_m\rightarrow
0}\hat{\Gamma}_{ij}^{(e-ph)}(\bk,\omega_n;\bq,\nu_m)\simeq\tau_0$
for any finite $\bq$, the electron-phonon interaction can be
suppressed in evaluating the vertex parts. The Coulomb
interaction, on the other hand, leads to Stoner-type
enhancement,\cite{Mor62} which is unaffected by the
superconducting transition (assuming the usual electron-phonon
pairing mechanism) and thus should cancel out from the ratio
$R_s/R_n$. Hence, we replace $\hat{\Gamma}_{ij}$ in Eq.
(\ref{strong_Kpm nu m}) with the unit matrix $\tau_0$ in computing
the ratio of the spin-lattice relaxation rates in the
superconducting and normal states. We note, however, that the
single particle energies $\xi_{i,\bk}$ are assumed to be
renormalized by the Coulomb interaction and that the
electron-phonon vertices entering various self-energy parts in
$\hat G_i(\bk,\omega_n)$ are Coulomb vertex corrected and Coulomb
renormalized as discussed in Ref. \onlinecite{Scalapino69}.

Next, one introduces the spectral representation for $\hat
G_i(\bk,\omega_n)$
\begin{equation}
\label{spectral1}
    \hat G_i(\bk,\omega_n)=\int\limits_{-\infty}^{+\infty}d\omega\frac{\hat
    A_i(\bk,\omega)}{i\omega_n-\omega},
\end{equation}
with
\begin{equation}
\label{spectral2}
    \hat A_i(\bk,\omega)=-\frac{1}{\pi}\im \hat G_i(\bk,\omega+i0^+),
\end{equation}
which allows one to calculate the Matsubara sums in Eq.
(\ref{strong_Kpm nu m}), followed by the analytical continuation
$i\nu_{m}\to\omega_0+i0^+$. In the limit $\omega_0\to 0$ we obtain
\begin{widetext}
\begin{eqnarray}
\label{strong_Kpm limit}
    &&\lim_{\omega_{0}\to 0}-\frac{1}{\pi}\frac{\im K(\omega_{0}+i0^+)}{\omega_{0}}
    =\frac{1}{\pi^2}\int\limits_{\bk_{1,2}}\int\limits_{-\infty}^{+\infty}d\omega
    \left(-\frac{\partial f}{\partial
    \omega}\right)\sum_{i,j}|u_{i,\bk_1}(\bo)|^2|u_{j,\bk_2}(\bo)|^2\nonumber\\
    &&\qquad\times\left[\im\frac{\omega
    Z_{i,\bk_1}(\omega)}{D_{i,\bk_1}(\omega)}
    \im\frac{\omega Z_{j,\bk_2}(\omega)}{D_{j,\bk_2}(\omega)}
    +\im\frac{\xi_{i,\bk_1}}{D_{i,\bk_1}(\omega)}
    \im\frac{\xi_{j,\bk_2}}{D_{j,\bk_2}(\omega)}
    +\im\frac{\phi_{i,\bk_1}(\omega)}{D_{i,\bk_1}(\omega)}
    \im\frac{\phi_{j,\bk_2}(\omega)}{D_{j,\bk_2}(\omega)}\right]
\end{eqnarray}
\end{widetext}
where
\begin{equation}
\label{denominator}
    D_{i,\bk}(\omega)=[\omega Z_{i,\bk}(\omega)]^2-\xi_{i,\bk}^2-\phi^2_{i,\bk}(\omega),
\end{equation}
and $Z_{i,\bk}(\omega)\equiv Z_{i,\bk}(\omega+i0^+)$,
$\phi_{i,\bk}(\omega)\equiv \phi_{i,\bk}(\omega+i0^+)$.

Next, we assume that $Z_{i,\bk}(\omega)$ and
$\phi_{i,\bk}(\omega)$ are isotropic, which seems to be a
reasonable assumption for MgB$_2$,\cite{Golub02} and use a weak
dependence of these functions on $\xi_{i,\bk}$ which is one of the
consequences of the Migdal's theorem. Hence, the $\bk$-dependence
of $Z_i$ and $\phi_i$ can be suppressed, and after defining the
local densities of states (\ref{N ab}), (\ref{M ab}), and (\ref{Nn
ab}), the momentum integrations in Eq. (\ref{strong_Kpm limit})
can be easily performed. The final result has the form
\begin{equation}
\label{strong_T1T final result}
    \frac{R_s}{R_n}=2\int\limits_0^{+\infty}d\omega
    \left(-\frac{\partial f}{\partial\omega}\right)
    \frac{N^2(\omega)+M^2(\omega)}{N_n^2},
\end{equation}
where
\begin{eqnarray}
\label{Ns ab strong}
    N(\omega)&=&\sum_iN_{n,i}\re\frac{\omega}{\sqrt{\omega^2-\Delta_i^2(\omega)}}, \\
\label{Ms ab strong}
    M(\omega)&=&\sum_iN_{n,i}\re\frac{\Delta_i(\omega)}{\sqrt{\omega^2-
    \Delta_i^2(\omega)}},
\end{eqnarray}
and $\Delta_i(\omega)=\phi_i(\omega)/Z_i(\omega)$ is the gap
function in band $i$. In arriving at (\ref{strong_T1T final
result}) we have used $\Delta_i(-\omega+i0^+)=
\Delta_i^{*}(\omega+i0^+)$ which follows directly from the
spectral representation (\ref{spectral1}). It is easy to see that
our result (\ref{strong_T1T final result}), (\ref{Ns ab strong}),
(\ref{Ms ab strong}) reduces to the one given by Fibich
\cite{Fibich65} in the case of a single isotropic band, and to
Eqs. (\ref{T1T final result}), (\ref{Ns convent}), (\ref{Ms
convent}) in the weak coupling limit, when the gap function does
not depend on $\omega$. Similar to the single-band case, the
presence of non-zero imaginary parts in $\Delta_i(\omega)$ leads
to the smearing out of the BCS square-root singularities in
$N(\omega)$ and $M(\omega)$.

\section{Application to M\lowercase{g}B$_2$}
\label{sec: MgB2}

For a quantitative application of the results of the previous
section to a particular compound, one needs to know both the
band-structure characteristics and the interaction parameters of
the Eliashberg theory. The only two-band superconductor for which
these are presently available is MgB$_2$.

Different contributions to the hyperfine interaction in MgB$_2$
were calculated using the local-density approximation in Refs.
\onlinecite{BAR01,PM01}. It was found that, while the relaxation
at the ${}^{25}$Mg nucleus is dominated by the Fermi contact
interaction, for the ${}^{11}$B nucleus it is the interaction with
the orbital part of the hyperfine field that makes the biggest
contribution. These predictions were subsequently found to be in
excellent agreement with experiments in the normal
state.\cite{Mali02,Gerash02,Papa02} To the best of our knowledge,
the experimental results on temperature dependence of $T_1^{-1}$
in the superconducting state of MgB$_2$ are available only for the
${}^{11}$B nucleus.\cite{Kote01,Kote02,Jung01,Baek02} Therefore
our theory, which should be applicable only to the relaxation rate
for the ${}^{25}$Mg nucleus in a clean sample, cannot be directly
verified by comparison with the existing experimental data. The
lack of data on $T_1^{-1}$ for the ${}^{25}$Mg nucleus is
presumably related to the small magnetic moment and a low natural
abundance of this nucleus as discussed in Ref.
\onlinecite{Mali02}. Nevertheless, the experiments performed in
Refs. \onlinecite{Mali02,Gerash02} indicate that it is possible in
principle to measure $^{25}R$ below the superconducting transition
temperature.

\begin{figure}
\includegraphics[angle=0,width=8cm]{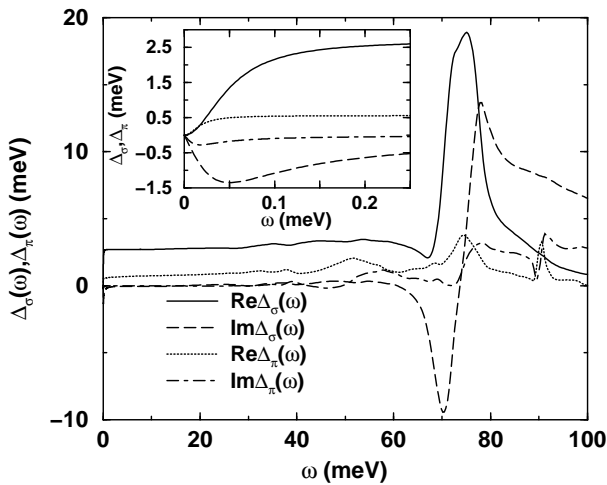}
\caption{The solutions for the real and imaginary parts of
$\Delta_{\sigma}(\omega)$ and $\Delta_{\pi}(\omega)$ in the entire
phonon energy range for MgB$_2$, at $T$ = 0.968$T_{c}$ . The inset
shows the solutions in the low energy range where the real parts
of the gaps are quadratic functions of $\omega$ and the imaginary
parts of the gaps are linear functions of $\omega$ at low enough
energy for $T>$0.} \label{fig:gaps}
\end{figure}

To calculate $R_s/R_n$ in the superconducting state of MgB$_2$ we
have solved the coupled Eliashberg equations with the realistic
interaction parameters for the isotropic two-band model,
\cite{Golub02} on the real frequency axis and at finite
temperature:
\begin{widetext}
\begin{eqnarray}
\label{Eli1}
    \Delta_{i}(\omega)Z_{i}(\omega)  =  \sum_j\int\limits_0^{\omega_c}d\omega'
    \re\frac{\Delta_j(\omega')}{\sqrt{{\omega'}^2-\Delta_j^2(\omega')}}
    \Big[f(-\omega')K_{+,ij}(\omega,\omega')-f(\omega')K_{+,ij}(\omega,-\omega')
    \Big. \nonumber \\
    \left.
    -\mu_{ij}^{*}(\omega_c)\tanh\frac{\omega'}{2T}
    +K_{+,ij}^{TP}(\omega,\omega')-K_{+,ij}^{TP}(\omega,-\omega')
    \right]\>,
\end{eqnarray}
\begin{eqnarray}
\label{Eli2}
    Z_{i}(\omega)=1-\frac{1}{\omega} \sum_j\int\limits_0^{+\infty}d\omega'
    \re\frac{\omega'}{\sqrt{{\omega'}^2-\Delta_j^2(\omega')}}
    \left[f(-\omega')K_{-,ij}(\omega,\omega')-f(\omega')K_{-,ij}(\omega,-\omega')
    \right. \nonumber \\
    \left.
    +K_{-,ij}^{TP}(\omega,\omega')-K_{-,ij}^{TP}(\omega,-\omega')
    \right]\>,
\end{eqnarray}
\end{widetext}
where
\begin{widetext}
\begin{eqnarray}
\label{kernel}
    K_{\pm,ij}(\omega,\omega') & =
    &\int\limits_0^{+\infty}d\Omega\;
    \alpha^{2}F_{ij}(\Omega)\left[\frac{1}{\omega'+\omega+\Omega+i0^{+}}\pm
    \frac{1}{\omega'-\omega+\Omega-i0^{+}}\right]\>, \\
\label{TPkernel}
    K_{\pm,ij}^{TP}(\omega,\omega') & =
    &\int\limits_0^{+\infty}d\Omega\;
    \frac{\alpha^{2}F_{ij}(\Omega)}{e^{\Omega/T}-1}
    \left[\frac{1}{\omega'+\omega+\Omega+i0^{+}}\pm
    \frac{1}{\omega'-\omega+\Omega-i0^{+}}\right]\>.
\end{eqnarray}
\end{widetext}
With a set of four electron-phonon coupling functions
$\alpha^{2}F_{ij}(\Omega)$, $i,j=\sigma,\pi$, calculated in Ref.
\onlinecite{Golub02}, and with a set of the Coulomb repulsion
parameters $\mu_{ij}^*(\omega_c)$ determined in Ref.
\onlinecite{Mitro04} to fit the experimental critical temperature
$T_{c}$, Eqs.~(\ref{Eli1},\ref{Eli2}) were solved for the complex
gap functions $\Delta_{\sigma}(\omega)$ and $\Delta_{\pi}(\omega)$
at a series of temperatures below $T_{c}$. A representative
solution near $T_{c}$ is shown in Fig. \ref{fig:gaps}
($T=0.968T_c$).

\begin{figure}
\includegraphics[angle=0,width=7.4cm]{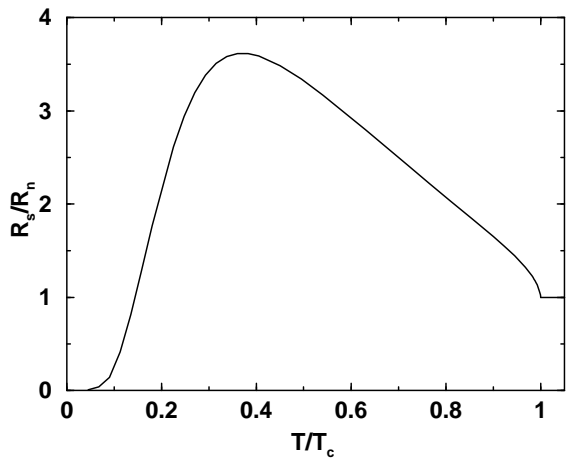}
\caption{The ratio $R_{s}/R_{n}$ as a function of the reduced
temperature $T/T_c$ in the case when the relaxation is dominated
by the lower-gap band. } \label{fig:rate_strong}
\end{figure}

The band structure calculations \cite{Bose05} indicate that the
contribution to the local density of states  at the Mg site from
the $\sigma$ band is much smaller than that from the $\pi$ band.
Therefore we can set $N_{\sigma}=0$ in the expressions for
$T_1^{-1}$ on the ${}^{25}$Mg nucleus. In
Fig.~\ref{fig:rate_strong} we show the temperature dependence of
$R_s/R_n$ obtained from the numerical solutions of the
strong-coupling gap equations, using Eqs. (\ref{strong_T1T final
result}, \ref{Ns ab strong},\ref{Ms ab strong}). At the lowest
temperatures, the relaxation rate is exponentially small, while at
$T\to T_c-0$, $R_{s}/R_{n}-1$ is proportional to
$(1-T/T_{c})^{0.5}$. The most prominent qualitative feature is a
shift of the Hebel-Slicher peak away from $T_c$ to a lower
temperature, at which the coherence factor from the lower gap in
the $\pi$-band makes the maximum contribution. The significant
increase in the peak's height can be attributed to a reduction of
the gap broadening due to the lifetime effects at lower
temperatures. This is in turn related to the fact that MgB$_2$ is
not a very-strong-coupling superconductor. If it were then one
could expect the Hebel-Slichter peak to be suppressed, similar to
the single-band case.\cite{AR91,AC91}

\section{Conclusions}

We calculated the NMR relaxation rate $T_1^{-1}$ in a singlet
two-band superconductor without spin-orbit coupling and
impurities, assuming that the relaxation of the nuclear spins is
dominated by the Fermi contact interaction with the band
electrons. Our main result is that there are important inter-band
contributions not related to any scattering processes, which
change the temperature dependence of the relaxation rate. In
particular, if there are unpaired sheets of the Fermi surface in a
superconductor with gap nodes, then in addition to the residual
relaxation rate at $T=0$, one should see unusual exponents in the
power-law behavior at low $T$. The observation of those exponents
could be a strong argument in favor of multi-band
superconductivity.

To illustrate the general theory, we calculated the relaxation
rates in the clean limit for (i) a two-dimensional $d$-wave
superconductor, using the BCS theory, and (ii) an isotropic
$s$-wave superconductor, for which a strong-coupling treatment is
required. In the latter case, we applied our model to the
${}^{25}$Mg nucleus in MgB$_2$, for which the relaxation is due to
the Fermi contact interaction and the parameters of the Eliashberg
theory are known. The predicted temperature dependence of the
relaxation rate is quite unusual and should be easily detectable
in experiments.

In order to expand the applicability of our theory, one should
include disorder, especially the interband scattering, which is a
pair-breaker in the multi-band superconductors. Although the
unconventional candidates for multi-band superconductivity, such
as CeCoIn$_5$, are in the clean limit, in general the impurity
effects might be significant. Also, our basic assumption that the
relaxation is controlled by the local fluctuations of the
Fermi-contact hyperfine field, can be violated in some cases, e.g.
for the ${}^{11}$B nucleus in MgB$_2$. Another possible
generalization would include the effects of the gap anisotropy
within the separate bands.\cite{Choi02} It is well known
\cite{Tink96} that the spread in gaps within a single band leads
to the suppression of the coherence peak in $R_{s}/R_{n}$ below
$T_c$. Finally, if the NMR measurements are done at a non-zero
magnetic field in the presence of vortices, then the inhomogeneity
in the order parameter in the mixed state strongly affects the
density of quasiparticle states and therefore the relaxation
rate.\cite{EPT66}

\acknowledgements

We thank S. Bose for helpful discussions and L. Taillefer for
informing us about the recent experiments in CeCoIn$_5$. This work
was supported by the Natural Sciences and Engineering Research
Council (NSERC) of Canada.

\appendix

\section{Derivation of Eq. (\ref{T1T general})}
\label{sec: Appendix A}

We assume that the dominant mechanism of the spin-lattice
relaxation is the interaction between the nuclear spin magnetic
moment $\hbar\gamma_n\bm{I}$ ($\gamma_n$ is the nuclear
gyromagnetic ratio) and the hyperfine field created at the nucleus
by the conduction electrons. The system Hamiltonian is
$H=H_e+H_n+H_{int}$, where $H_e$ describes the electron subsystem,
$H_n=-\hbar\gamma_n\bm{I}\bm{H}$ is the Zeeman coupling of the
nuclear spin with the external field $\bm{H}$, and
\begin{equation}
\label{H int hf}
    H_{int}=-\hbar\gamma_n\bm{I}\bm{h}
\end{equation}
is the hyperfine interaction. For $I=1/2$, we have two nuclear
spin states $I_z=\pm 1/2$ with the energies
$E_{I_z}=-\hbar\omega_0 I_z$, where $\omega_0=\gamma_nH$ is the
NMR frequency and the spin quantization axis is chosen along
$\bm{H}$. The hyperfine field $\bm{h}$ can be represented as a sum
of the Fermi contact, the orbital, and the spin-dipolar
contributions \cite{Slichter90}. Their relative importance depends
on the electronic structure and therefore varies for different
systems. For example, if the Fermi contact interaction is
dominant, then $\bm{h}=-(8\pi/3)\hbar\gamma_e\bm{S}(\bo)$, where
$\gamma_e$ is the electron gyromagnetic ratio and
$\bm{S}(\br)=(1/2)\bm{\sigma}_{\alpha\beta}\psi_\alpha^\dagger(\br)\psi_\beta(\br)$
is the electron spin density at $\br=\bo$. The derivation below
does not rely on any particular expression for the hyperfine
field.

According to Ref. \onlinecite{Slichter90}, the relaxation rate for
a spin-$1/2$ nucleus is given by
\begin{equation}
\label{T1 Ws}
    \frac{1}{T_1}=W_{+-}+W_{-+},
\end{equation}
where $W_{+-}$ and $W_{-+}$ are the transition probabilities per
unit time, from $I_z=+1/2$ to $I_z=-1/2$ and from $I_z=-1/2$ to
$I_z=+1/2$, respectively. The hyperfine interaction is usually
small, which makes it possible to use the lowest-order
perturbation theory to calculate $W_{+-}$ and $W_{-+}$. The states
of the whole system at zero hyperfine coupling can be represented
as $|I\rangle=|i,I_z\rangle$, where $i$ labels the exact (in
general, many-particle) eigenstates of $H_e$, with energies $E_i$.
When $J\neq 0$, then the transition probability per unit time from
an initial state $|I\rangle$ of energy ${\cal E}_I$ to a final
state $|F\rangle$ of energy ${\cal E}_F$ can be found using the
Golden Rule:
\begin{equation}
\label{Golden Rule}
    w_{|I\rangle\to|F\rangle}=\frac{2\pi}{\hbar}|\langle
    I|H_{int}|F\rangle|^2\delta({\cal E}_I-{\cal E}_F).
\end{equation}
The transition rates for the nuclear spin are calculated in the
usual fashion by averaging over the initial and summing over the
final electron states.

For $W_{+-}$, we have $|I\rangle=|i,+1/2\rangle$, ${\cal
E}_I=E_i-\hbar\omega_0/2$ and $|F\rangle=|f,-1/2\rangle$, ${\cal
E}_F=E_f+\hbar\omega_0/2$. Then
\begin{equation}
\label{W+-}
    W_{+-}=\sum_i\rho_{e,i}\sum_f
    w_{|i,+1/2\rangle\to|f,-1/2\rangle},
\end{equation}
where $\rho_e=e^{-\beta H_e}/\mathrm{Tr}\,e^{-\beta H_e}$ is the
density matrix of the electron subsystem. Inserting here the
expressions (\ref{Golden Rule}) and (\ref{H int hf}) and
representing $\bm{I}\bm{h}=I_zh_z+(I_+h_-+I_-h_+)/2$, where
$I_\pm=I_x\pm iI_y$ and $h_\pm=h_x\pm ih_y$, we find that only the
$I_+h_-$ term makes a non-zero contribution. Therefore,
$$
    W_{+-}=\frac{\pi\hbar\gamma_n^2}{2}\sum_{i,f}\rho_{e,i}|\langle i|h_-
    |f\rangle|^2\delta(E_i-E_f-\hbar\omega_0).
$$
This expression can be simplified by using the identity
$$
    \delta(E_i-E_f-\hbar\omega_0)=\int_{-\infty}^{\infty}
    \frac{dt}{2\pi\hbar}e^{i(E_i-E_f-\hbar\omega_0)t/\hbar}
$$
and the fact that $h_-^\dagger=h_+$, which allow us to write
\begin{eqnarray*}
    &&|\langle i|h_-|f\rangle|^2e^{i(E_i-E_f)t/\hbar}\\
    &&=\langle i|e^{iE_it/\hbar}h_-e^{-iE_ft/\hbar}|f\rangle\langle
    f|h_+|i\rangle\\
    &&=\langle i|h_-(t)|f\rangle\langle f|h_+(0)|i\rangle,
\end{eqnarray*}
where $h_\pm(t)=e^{iH_et/\hbar}h_\pm e^{-iH_et/\hbar}$. Now the
sum over the final states can be calculated, and we finally have
\begin{equation}
\label{W+- final}
    W_{+-}=\frac{\gamma_n^2}{4}\int_{-\infty}^{\infty}dt\;e^{-i\omega_0t}
    \langle h_-(t)h_+(0)\rangle.
\end{equation}
The angular brackets here stand for the thermal averaging with
respect to the electron density matrix $\rho_e$. Similarly, we
obtain
\begin{equation}
\label{W-+ final}
    W_{-+}=\frac{\gamma_n^2}{4}\int_{-\infty}^{\infty}dt\;e^{i\omega_0t}
    \langle h_+(t)h_-(0)\rangle.
\end{equation}

Combining Eqs. (\ref{W+- final}) and (\ref{W-+ final}), we have
\begin{equation}
\label{T1 anticomm}
    \frac{1}{T_1}=\frac{\gamma_n^2}{4}\int_{-\infty}^{\infty}dt\;e^{i\omega_0t}
    \langle\{h_+(t),h_-(0)\}\rangle.
\end{equation}
The integral on the right-hand side here can be expressed in terms
of the Fourier transform of the retarded correlator of the
hyperfine fields
$K_{hh}^R(t)=-i\langle[h_+(t),h_-(0)]\rangle\theta(t)$, giving
\begin{eqnarray}
\label{T1 general}
    \frac{1}{T_1}&=&-\frac{\gamma_n^2}{4\pi}\coth\left(\frac{\hbar\omega_0}{2k_BT}\right)
    \im K_{hh}^R(\omega_0)\nonumber\\
    &\simeq&-\frac{\gamma_n^2}{2\pi}\frac{k_BT}{\hbar\omega_0}
    \im K_{hh}^R(\omega_0).
\end{eqnarray}
Here we used the fact that in a typical experiment the condition
$\hbar\omega_0\ll k_BT$ is always satisfied [we also note that
since $W_{+-}/W_{-+}=e^{-\beta\hbar\omega_0}\simeq 1$ due to the
detailed balance in the thermal equilibrium, one could use
$T_1^{-1}\simeq 2W_{+-}$ instead of (\ref{T1 Ws})]. Keeping only
the Fermi contact term in the hyperfine interaction (\ref{H int
hf}), we finally arrive at Eq. (\ref{T1T general}) with
$J=(8\pi/3)\gamma_n\gamma_e$.

\end{document}